\documentclass[sigconf]{acmart} 
\UseRawInputEncoding
\usepackage{amsthm}
\usepackage{amsfonts} 
\usepackage{url}          
\usepackage{subfigure}
\usepackage{ifthen}
\usepackage{color, xcolor}
\hypersetup{colorlinks=true, linkcolor=blue, citecolor=blue, anchorcolor=blue, urlcolor=blue}
\usepackage{thm-restate}
\usepackage{algorithm}
\usepackage{algorithmic}
\usepackage{enumitem}
\usepackage{multirow}
\usepackage{array}
\usepackage{microtype}

\AtBeginDocument{%
 }

\copyrightyear{2023}
\acmYear{2023}
\setcopyright{rightsretained}
\acmConference[KDD '23]{Proceedings of the 29th ACM SIGKDD Conference on Knowledge Discovery and Data Mining}{August 6--10, 2023}{Long Beach, CA, USA}
\acmBooktitle{Proceedings of the 29th ACM SIGKDD Conference on Knowledge Discovery and Data Mining (KDD '23), August 6--10, 2023, Long Beach, CA, USA}
\acmDOI{10.1145/3580305.3599873}
\acmISBN{979-8-4007-0103-0/23/08}

\renewcommand\footnotetextcopyrightpermission[1]{}  
\newcommand{\compilehidecomments}{false}
\ifthenelse{ \equal{\compilehidecomments}{true} }{%
	\newcommand{\wei}[1]{}
	\newcommand{\sheng}[1]{}
	\newcommand{\hao}[1]{}
	\newcommand{\zhou}[1]{}
        \newcommand{\zhang}[1]{}
}{
	\newcommand{\wei}[1]{{\color{blue!50!black}  [\text{Wei:} #1]}}
	\newcommand{\sheng}[1]{{\color{red!70!black} [\text{Sheng:} #1]}}
	\newcommand{\hao}[1]{{\color{green!90!black} [\text{Hao:} #1]}}
	\newcommand{\zhou}[1]{{\color{yellow!90!black} [\text{zhou:} #1]}}
        \newcommand{\zhang}[1]{{\color{purple!90!black} [\text{zhang:} #1]}}
}

\makeatletter
\gdef\@copyrightpermission{
  \begin{minipage}{0.3\columnwidth}
   \href{https://creativecommons.org/licenses/by/4.0/}{\includegraphics[width=0.90\textwidth]{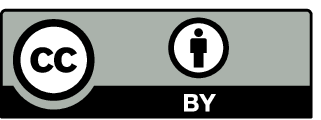}}
  \end{minipage}\hfill
  \begin{minipage}{0.7\columnwidth}
   \href{https://creativecommons.org/licenses/by/4.0/}{This work is licensed under a Creative Commons Attribution International 4.0 License.}
  \end{minipage}
  \vspace{5pt}
}
\makeatother

\begin{document}

\title{MUSER: A MUlti-Step Evidence Retrieval Enhancement Framework for Fake News Detection}

\author{Hao Liao}
\affiliation{%
  \institution{Shenzhen University}
  \city{Shenzhen}
  \country{China}
}
\email{haoliao@szu.edu.cn}

\author{Jiahao Peng}
\affiliation{%
  \institution{Shenzhen University}
  \city{Shenzhen}
  \country{China}
}
\email{2070276145@email.szu.edu.cn}

\author{Zhanyi Huang}
\affiliation{%
  \institution{Shenzhen University}
  \city{Shenzhen}
  \country{China}
}
\email{huangzhanyi2020@email.szu.edu.cn}

\author{Wei Zhang}
\affiliation{%
  \institution{Shenzhen University}
  \city{Shenzhen}
  \country{China}
}
\email{2210275010@email.szu.edu.cn}

\author{Guanghua Li}
\affiliation{%
  \institution{Shenzhen University}
  \city{Shenzhen}
  \country{China}
}
\email{2210275050@email.szu.edu.cn}

\author{Kai Shu}
\authornotemark[1]
\affiliation{%
  \institution{Illinois Institute of Technology}
  \city{Chicago}
  \country{USA}}
\email{kshu@iit.edu}

\author{Xing Xie}
\authornotemark[1]
\affiliation{%
  \institution{Microsoft Research Asia}
  \city{Beijing}
  \country{China}
}
\email{xingx@microsoft.com}


\renewcommand{\shortauthors}{Hao Liao and Jiahao Peng et al.}

\begin{abstract}

The ease of spreading false information online enables individuals with malicious intent to manipulate public opinion and destabilize social stability. Recently, fake news detection based on evidence retrieval has gained popularity in an effort to identify fake news reliably and reduce its impact. Evidence retrieval-based methods can improve the reliability of fake news detection by computing the textual consistency between the evidence and the claim in the news.
In this paper, we propose a framework for fake news detection based on \underline{\textbf{MU}}lti-\underline{\textbf{S}}tep \underline{\textbf{E}}vidence \underline{\textbf{R}}etrieval enhancement (MUSER), which simulates the steps of human beings in the process of reading news, summarizing, consulting materials, and inferring whether the news is true or fake. Our model can explicitly model dependencies among multiple pieces of evidence, and perform multi-step associations for the evidence required for news verification through multi-step retrieval. In addition, our model is able to automatically collect existing evidence through paragraph retrieval and key evidence selection, which can save the tedious process of manual evidence collection. 
We conducted extensive experiments on real-world datasets in different languages, and the results demonstrate that our proposed model outperforms state-of-the-art baseline methods for detecting fake news by at least 3\% in F1-Macro and 4\% in F1-Micro. Furthermore, it provides interpretable evidence for end users.

\end{abstract}



\maketitle

\begin{figure}[t]
  \centering
  \includegraphics[width=0.95\linewidth]{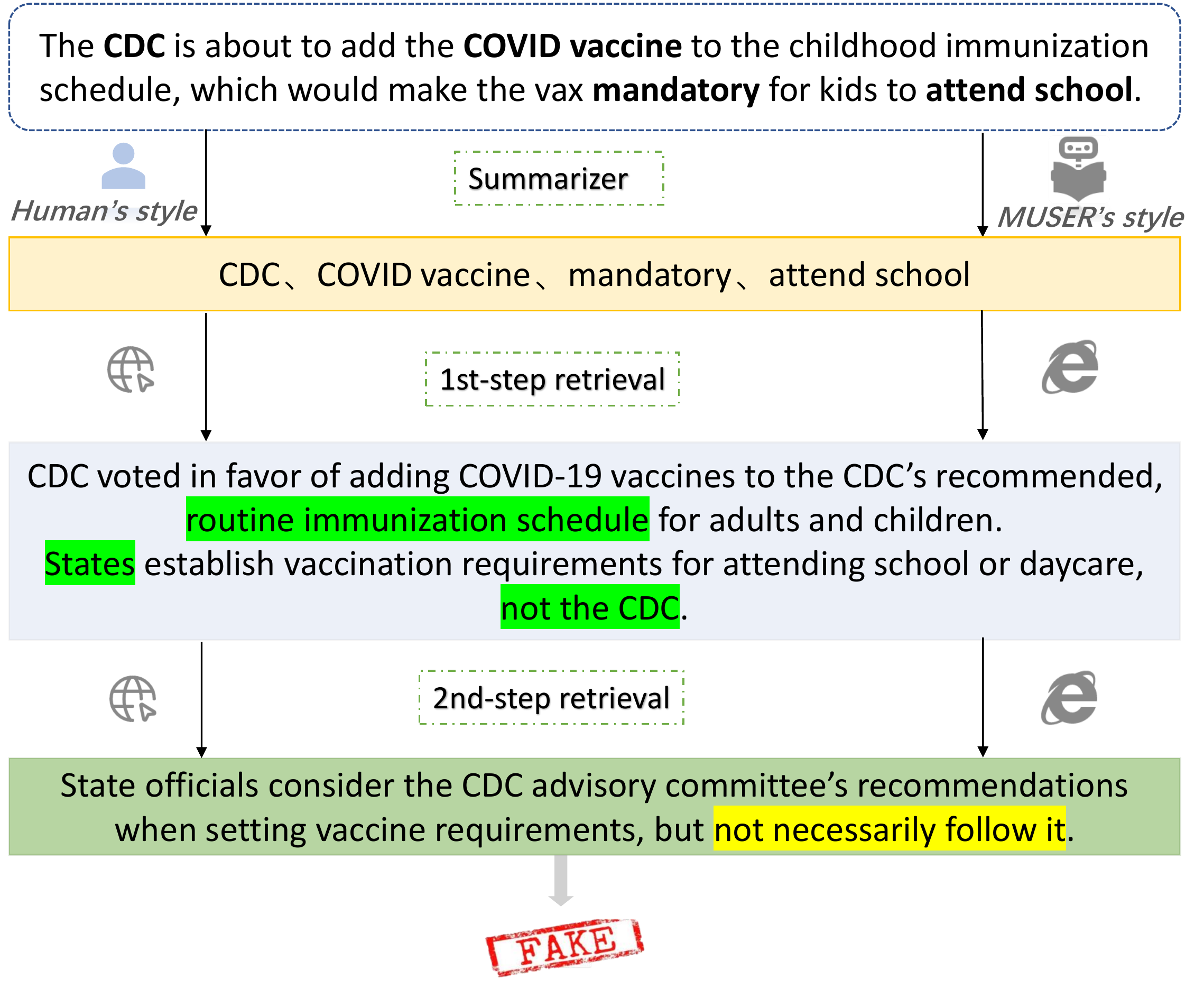}
  \caption{ A motivating example of MUSER model. Our model simulates a human evaluating news through three steps:
(1) Summarization of the key information,
(2) Retrieval and evaluation of relevant evidence: the model assesses the sufficiency and quality of the evidence, determining if additional inquiries are necessary,
(3) Conclusion regarding the truthfulness of the news based on the gathered evidence.
}
  \label{example1}
\end{figure}

\section{Introduction}

The explosive growth of fake news has exerted serious negative consequences across society affecting areas such as politics, the economy, and public health~\cite{2017smfn}. This phenomenon is characterized by the dissemination of sensationalized and alarmist content, which caters to the mindset of netizens and is easily exploited by the "headline party" ~\cite{Grinberg19}. To garner more attention, individuals are prone to share news articles or retweet tweets featuring captivating headlines without conducting a diligent evaluation. Consequently, this has facilitated the rapid dissemination of fake news through social media platforms, outpacing the circulation of authentic news. ~\cite{PopatWWW2017}. An overwhelming amount of fake news on social media has made it difficult for individuals to distinguish truth from falsehood, thereby posing a substantial threat to societal stability~\cite{yang2019unsupervised, ZHANG2020102025}. In light of these challenges, the emerging automated fake news detection has drawn widespread attention.


Generally, the detrimental effects of fake news tend to exacerbate over time. To mitigate the ramifications of fake news dissemination, it is important to promptly identify them on social platforms. Meanwhile, fake news detection can help netizens improve their ability to distinguish between true and fake news, thereby fostering the well-being and sustainability of social networks. Various efforts have been made by websites and social media platforms to combat fake news, such as Meta's encouragement for users to report untrustworthy posts and Sina Weibo's provision of a channel for debunking rumors~\cite{SONG2021102712TGNN}. Besides, fact-checking sites like FactCheck\footnote{\url{https://www.factcheck.org/}}, PolitiFact\footnote{\url{https://www.politifact.com/}} and Full Fact\footnote{\url{https://fullfact.org/}} 
have also begun to hire professionals to conduct fact-checking. However, the diversity and complexity of the increasing volume of news data make manual verification a time-consuming and unscalable process. 


To tackle this problem, data mining and machine learning techniques were introduced to detect fake news~\cite{shu2017kdden,cha2020detecting}. 
Intuitively, the task of fake news detection can be framed as a binary classification problem. These methods commonly employ supervised learning techniques, utilizing textual features such as sentence semantics and news entities, to distinguish between genuine and fabricated news articles~\cite{GOLDANI2021106991, ZhangAndFang2019,potthast-etal-2018-stylometric}.
Though effective, these content-based methods exhibit some limitations, as fake news often resembles real news in textual features and lacks important information, such as social context~\cite{gupta2014tweetcred}. To overcome these limitations, multi-modal fake news detection frameworks have been proposed, which consider social context by analyzing news propagation patterns on social media, such as retweet relationship networks~\cite{Liu_Wu_2018, lu-li-2020-gcan}, and user-friend relationships~\cite{DAVOUDI2022116635, nguyen2020fang}. Fake news can spread rapidly and become difficult to control once it has reached a wide audience~\cite{spreadofnews2018Science}. Methods based on social context information require a substantial amount of social context information, which may not curb the dissemination of fake news in a timely manner. In addition to the temporal delay issue of detection, methods based on social context face the challenge of user privacy preservation. Therefore, recent research endeavors have increasingly focused on evidence-based verification techniques as a means to detect fake news. These methods perceive fake news detection as an inferential process, wherein external evidence is employed to scrutinize the veracity of the claims presented in news articles. By extracting and incorporating relevant information from the given evidence for claim verification, these methods aim to improve the interpretability of fake news detection. Notably, recent studies have showcased promising outcomes regarding the effectiveness of these approaches.~\cite{2019-han, 2018-declare,2021-mac,2022-get}. 

 
Despite substantial advancements over the years, fake news detection still confronts numerous challenges. Evidence-based detection methods suffer from the assumption that evidence is easily accessible, ignoring the large amount of manual effort required for evidence collection. Furthermore, prior work has inadequately explored complex, long-range semantic dependencies in evidence, neglecting the intricate relationships between information.
 
 
 
Inspired by brain science~\cite{brain_sciencePNAS}, we propose a fake news inference framework \textbf{MU}lti-\textbf{S}tep \textbf{E}vidence \textbf{R}etrieval (MUSER). The cognitive processes involved in human news consumption  typically involve three steps ~\cite{brain_science1} as shown in Figure \ref{example1}: First, a summary of the key findings or claims in the text is made. Second, supporting evidence for the claims is located and evaluated for quality, which may include sources such as website data, official experiments, or research. Finally, conclusions are drawn based on the evaluated evidence. By following these steps, it is possible to ascertain the sources of information, the evidence used, evidence quality, and limitations, thus helping readers to make informed judgments about the validity of the information. 
MUSER\footnote{Code is available at \url{https://github.com/Complex-data/MUSER/}} automatically retrieves existing evidence from Wikipedia through paragraph retrieval and key evidence selection, eliminating the need for manual evidence collection. Pieces of evidence needed for news verification are correlated through multi-step retrieval. Furthermore, our model can perform early detection without relying on social context information and provides reasons for the authenticity of the news through retrieved evidence. Although social media can provide external information for early fake news detection, there are two drawbacks - privacy concerns related to user comments and the presence of noisy information among user posts. Our main contributions can be summarized as follows:


 \begin{itemize}[leftmargin=*]
     \item We propose an automatic fact-checking framework for fake news detection that is based on multi-step evidence retrieval. Our framework can explicitly model dependencies among multiple pieces of evidence and retrieves the evidence necessary for news verification through multi-step retrieval. The framework simulates the searching behavior of people when verifying news content on the Internet, making it possible to narrow the gap between computers and human experts in fake news detection.
     \item The implementation of our proposed model includes three core modules: text summarization, multi-step retrieval, and text reasoning. In the multi-step retrieval module, we employ the method of key evidence selection to control the number of hops, realizing adaptive retrieval step control.
     \item We conduct extensive experiments on three real-world datasets 
     , and the results demonstrate the effectiveness of our model in terms of improved interpretability and good performance when compared with state-of-the-art models.
 \end{itemize}

\section{Related Work}
\subsection{Fake News Detection}
In recent years, researchers have collaborated with the news ecosystem to better define and characterize fake news through news content and social feedback from web users. We briefly introduce related work from the following aspects: 1) content-based; 2) social context-based; 3) evidence-based.

\textbf{Content-based:} Content-based methods detect fake news by exploiting news text, writing style, or external knowledge about news entities. Some works detect fake news by extracting news text features, e.g., 
n-gram distribution and/or utilize Linguistic Inquiry and Word Count (LIWC)~\cite{pennebaker2015development} features and sentence relationships based on Rhetorical Structure Theory (RST)~\cite{RST2015}. The stylistic feature-based approach distinguishes between real and fake news by capturing the specific writing style and emotion usually present in the textual content of fake news~\cite{2020styleoffakenews}. KAN~\cite{KAN2021} directly evaluates the authenticity of news by comparing news knowledge with knowledge entities in the knowledge graph. Content-based methods are often used in the early detection of fake news to curb the spread of rumors in the early stages of news dissemination.

\textbf{Social context-based:} Social media plays an important role in detecting fake news~\cite{Zubiaga2018}. It has been used to improve the performance of fake news detection by integrating contextual information on social platforms, such as user characteristics, comments, and positions~\cite{shu2019defend}. Methods based on communication structure rely on the assumption that the communication structure of real news and fake news is quite different~\cite{spreadofnews2018Science}. Network structure-based methods extract network features by constructing specific networks, such as user interaction networks, user social structures, participation patterns, and news dissemination networks~\cite{jin2022towards, PSIN2022, nguyen2020fang,wang2022kddsubgraph}.

\textbf{Evidence-based:} The semantic similarity (conflict) in the claim-evidence pairs can be used to determine the veracity of the news by searching Wikipedia or fact-checking websites according to the claims in the news. Early research approaches employ sequence models to embed semantics and apply attention mechanisms to capture claim-evidence semantic relations. For example, DeClar\cite{2018-declare} uses BiLSTM to embed the semantics of the evidence and calculates the evidence score through the attention interaction mechanism. MAC\cite{2021-mac} proposes a multi-level multi-head attention network combining word attention and evidence attention to detect fake news. GET\cite{2022-get} models the claims and evidence as graph-structured data, proposing a unified evidence-graph-based fake news detection method for the first time. Evidence-checking-based methods can reveal false parts of claims, provide users with evidence that news is true or fake, and improve the interpretability of fake news detection. Though effective, the above methods all assume that the evidence declared in the news already exists. However, the collection and arrangement of evidence in the actual process often require a lot of manual operations.

Different from the aforementioned studies, we propose a fake news inference framework augmented by multi-step evidence retrieval. Our model can automatically retrieve existing evidence through Wikipedia, conduct evidence collection, and capture dependencies among evidence through multi-step retrieval. 

\subsection{Retrieval Enhancement}
Recent work has shown that retrieving additional information can improve the performance of various downstream tasks~\cite{Retrieval-Augmented}. Such tasks include open-domain question answering, fact-checking, fact completion, long-form question answering, Wikipedia article generation, and dialogue. In the classic and simplest form of fact-checking, with claims as query conditions, the $k$ relevant passages $K_S = \{P_1, P_2, \ldots, P_{|K_S|}\} $ needed to verify the claims are obtained. Evidence may be contained within a paragraph, or even within a sentence. Retrieve multiple relevant passages $P_i \in K_S$ by a given query Q, and let the reading comprehension model extract the answer from $P_i$~\cite{guo-etal-2022-survey,kruengkrai-etal-2021-multi}.
These studies all used a single-step search. Contrary to the case of single-step retrieval, evidence for some types of queries cannot be obtained through a single retrieval and requires multiple iterative queries. The ability to retrieve information with multiple iterations is known in the literature as multi-step retrieval~\cite{feldman-el-yaniv-2019-multi}. In multi-step retrieval, evidence may need to be obtained with additional information from a previous search, which might otherwise be interpreted as not being fully relevant to the question and no evidence could be found.
We extend the capability of multi-step retrieval to fake news claim verification, querying relevant evidence passages in an iterative retrieval manner.

\subsection{Natural Language Inference}
Given a statement and selected evidence sentences, the task of NLI is to predict their relation labels $y$. The advent of large annotated datasets, such as SNLI~\cite{2015-SNLI}, CreditAssess~\cite{2016CreditAssess}, FEVER~\cite{thorne-etal-2018-fever}, has facilitated the development of many different neural NLI models, facilitating model development for this task~\cite{parikh-etal-2016-NLI1, NLI2}. The fact verification task related to natural language inference aims to classify a pair of claims and evidence extracted from Wikipedia into three categories: entailment, contradiction, or neutrality. NSMN~\cite{NLI2} uses a connected system of three homogeneous neural semantic matching models that jointly perform document retrieval, sentence selection, and claim verification for fact extraction and verification. Soleimani et al.~\cite{Soleimani} retrieve and validate claims using a BERT~\cite{bert2019} model. With the popularity of graph neural networks, graph-based models are also used for semantic reasoning. EVIN~\cite{ma2019sentence-evin} proposes an evidence reasoning network, which extracts core semantic conflicts of claims as evidence to explain verification results. Our work differs from prior research in that we focus on classifying news claims as true or false on a comprehensive examination of relevant evidence.

\section{Problem Statement}
In this section, we first define the problem of fake news detection based on evidence retrieval enhancement.
We draw a parallel between the detection of fake news and the process by which human beings verify the authenticity of a news article. First, we read the news content and summarize the key information expressed in the news (content summary), then query the evidence in multiple steps based on the summary (multi-step retrieval), and finally infer the authenticity of the news (i.e., Natural Language Inference).
So our problem is defined as follows: the input is only news text $A$, and then the news key statement $C$ is obtained through the text summarization module. Retrieve relevant passages in Wikipedia through $C$ to get $P = \{P_1, P_2, P_3, \ldots \}$, and then perform evidence extraction to obtain $E=\{e_1, e_2, e_3, \ldots\}$. 
The output is the predicted probability of news authenticity $\hat{y}= f(C,E)$, where $f$ is the natural language inference verification model. And $y \in \{0,1\}$ represents the binary classification labels. In this context, $y=0$ corresponds to fake news, while $y=1$ corresponds to true news. 

\begin{figure*}[t]
  \includegraphics[width=0.95\textwidth]{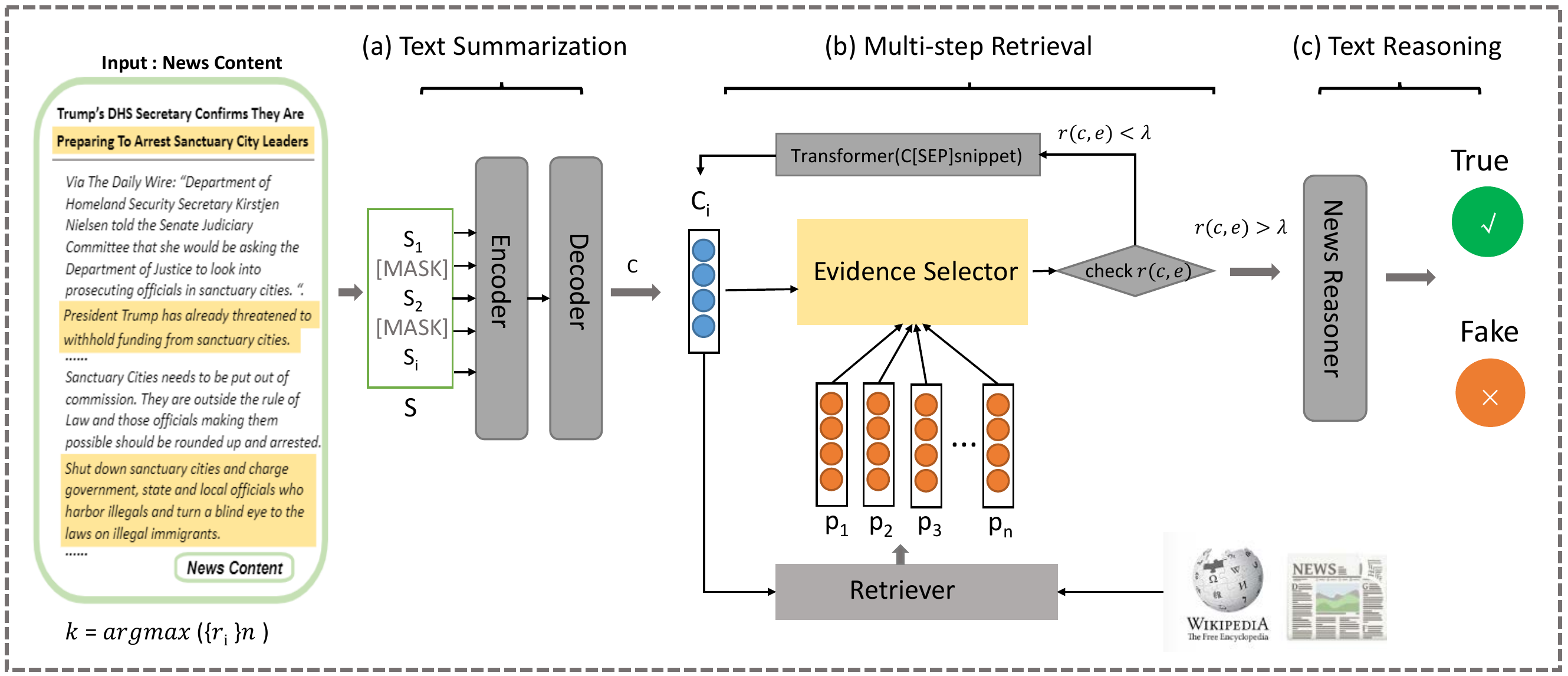}
  \caption{ Our framework unfolds in three steps: (a) Summarization of the initial news text to obtain the key statement $C$, corresponding to the human process of summarizing key information, (b) Evidence finding through multi-step retrieval, corresponding to the human process of querying external relevant information based on the news claim. The retriever sends the first $k$ paragraphs to the evidence selector, which evaluates whether the evidence meets the requirements. The correlation coefficient between $C$ and evidence snippets is represented by $r(c,e)$, and a settable correlation score threshold, $\lambda$, is used to judge the quality of the evidence, and (c) The textual reasoner infers the consistency of evidence and claims, corresponding to the human process of judging news based on evidence.}
  \Description{xxx}
  \label{MUSER}
\end{figure*}

\section{The Proposed Model}
In this section, we propose a framework for fake news detection based on \textbf{MU}lti-\textbf{S}tep \textbf{E}vidence \textbf{R}etrieval augmentation(MUSER). Figure \ref{MUSER} illustrates the overall architecture of MUSER. Our model mainly consists of three modules:

\textbf{Part 1: Text summarization module: }
Simulating the human behavior of reading news and summarizing key news information, the proposed module extracts the key information in the news and filters out the interference of redundant or unimportant information in the news.

\textbf{Part 2: Multi-step retrieval module: }
Simulating the behavior of humans querying external relevant information in response to news statements, we incorporate a retrieval module into our model. To handle situations where the initially retrieved paragraph may not contain the answer, we adopt a multi-step iterative retrieval method. This process starts by updating the query vector based on the key information and the current query vector. The retriever module then uses this updated query vector for re-retrieval, enabling a deeper exploration of relevant evidence. 

\textbf{Part 3: Text reasoning module: }
 Simulating the behavior of humans to judge true or fake news based on the supplementary information queried, this module can extract semantic links between news claims and evidence, and then classify news into two categories: true news and fake news.
Through the method of evidence retrieval enhancement, the interpretability of fake news detection is improved, thus mitigating the labor-intensive process of manual evidence extraction.

\subsection{Text Summarization Module}

Naturally, when reading a news article, individuals have a tendency to summarize the key content conveyed within. In order to simulate the ability of humans to summarize news information, we first pre-train a text summarization module. The purpose of this module is to extract the key information in the news and extract the statements worth checking.
Although pre-trained language models, such as BERT~\cite{bert2019} and UniLM~\cite{dong2019unified}, have achieved remarkable results in NLP scenarios, the word and subword mask language models used in the models may not be suitable for generative text summarization tasks. The reason is that the summarization task requires a coarser-grained semantic understanding, such as sentence and paragraph semantic level understanding, for an effective summary generation.

Inspired by the recent success in masking words and continuous spans, we pre-train a transformer-based encoder-decoder model on a large text corpus for news summarization generation~\cite{textsum}. To leverage a large text corpus for pre-training, we design a sequence-to-sequence self-supervised objective without abstract summarization. We mask sentences from news text and generate an output sequence from the remaining sentences for extracting news summaries. To enhance the relevance of the generated summaries, we select sentences that are deemed important or central to the news.

A piece of news $A$ contains multiple sentences, that is, $A = \{s_i\}^N_i$, where $N$ is the number of sentences. We select the set $S$ of $m$ sentences with the highest scores based on importance. As a proxy for importance, we compute ROUGE1-F1~\cite{lin-2004-rouge} between the sentence and the rest of the news.
\begin{equation}
  r_i=rouge(S \cup s_i,A  \verb|\| \{S \cup s_i\}),\quad\forall i,\, s_i \not\in S
\end{equation}
$A\setminus \{S \cup s_i\}$ represents the remaining sentences, and $S$ is initially an empty set. Then select important sentences according to the importance score $r_i$:
\begin{equation}
 k=argmax(\{r_i\}_n)
\end{equation}
\begin{equation}
 S = S \cup s_k
\end{equation}

The corresponding position of each selected sentence is replaced by a mask token [MASK] to inform the model.
Making $m$ selections, in the end, we select the masked $m$ sentences from the document and concatenate the sentences into a pseudo-summary.
the module then generates an output sequence from the remaining sentences, producing the masked sentences. We pre-train the model on the open source news dataset ~\footnote{\href{http://atp-modelzoo-sh.oss-cn-shanghai.aliyuncs.com/release/tutorials/generation/en\_train.tsv}{Engilsh: http://atp-modelzoo-sh.oss-cn-shanghai.aliyuncs.com/release/tutorials/gener ation/en\_train.tsv}\\ \href{http://atp-modelzoo-sh.oss-cn-shanghai.aliyuncs.com/release/tutorials/generation/cn\_train.tsv}{Chinese: http://atp-modelzoo-sh.oss-cn-shanghai.aliyuncs.com/release/tutorials/gener ation/cn\_train.tsv}} to achieve a better summary generation results.
The Mask sentences ratio (MSR) which refers to the ratio of the number of selected gap sentences to the total number of sentences in the document, is an important hyperparameter, similar to the mask rate in other works ~\cite{textsum}.
A low MSR reduces the difficulty and computational efficiency of pre-training. On the other hand, masking a large number of sentences at high MSR loses the contextual information necessary for guidance generation.
In our experiments, we found an MSR of 30\% to be effective.

\subsection{Multi-step Retrieval Module}
The purpose of this module is to perform retrieval enhancement based on the key information in the news extracted in the previous step, which is similar to humans looking up data, and finding supplementary information to assist in the identification of true and fake news.
Single-step retrieval may lead to insufficient auxiliary information retrieved. Therefore, we adopt a multi-step iterative retrieval method to improve information sufficiency~\cite{feldman-el-yaniv-2019-multi}. Through iterative retrieval and supplementation, relevant information can be extracted more comprehensively, so as to better assist in judging the authenticity of news.
When implementing this module, it is important to consider how to effectively extract the retrieved key information and how to maintain the sufficiency of information during the multi-step iterative retrieval process.

The multi-step retrieval problem we attempt to address is divided into three steps. In the first step, the news statement $C$ is used to retrieve the relevant paragraph $P$ from the Wikipedia corpus. The second step is to extract evidence from the retrieved long paragraphs and extract the key evidence of the paragraphs. Finally, in the case where no evidence is found in the retrieved paragraphs, the information retrieved in this step is fused with statement $C$ to generate a new statement for the retrieval iteration. The search terminates when evidence is found in the retrieved passages.

\textbf{Paragraphs retrieval}: Paragraphs retrieval is the selection of Paragraphs on Wikipedia that are relevant to a given statement. The paragraph retrieval module is based on BERT~\cite{bert2019} and creates dense vectors for paragraphs by computing their average token embedding. The relevance of paragraph $p$ to statement $c$ is given by their dot product:
\begin{equation}
 r(c,p)= \varphi(c)^T \varphi(p)
\end{equation}
$\varphi(\cdot)$ is an embedding function used to map paragraphs and statements to a dense vector. Dot product search can use the approximate nearest neighbor index implemented by the FAISS library to improve search efficiency ~\cite{johnson2019billionfaiss}. For the embedding function $\varphi(\cdot)$, we use the average token embedding of the BERT-base language model \cite{reimers-gurevych-2019-sentence}, which has been fine-tuned on several tasks:
\begin{equation}
\varphi(p) = \frac{1}{p}\sum_{i=1}^{|p|}BERT(p,i)
\end{equation}
where $BERT(p, i)$ is the embedding of the $i$-th token in paragraph $p$, and $|p|$ is the number of tokens in $p$.

\textbf{Key evidence selection}: Key evidence selection is to extract evidence-related key sentences from the retrieved relevant passages. Similar to paragraph retrieval, sentence selection can also be perceived as a semantic matching task, wherein each sentence within a paragraph is compared to a given statement query to identify the most plausible evidence interval. 
Since the search space has been reduced to a controllable size via the paragraph retrieval in the previous step, we can directly traverse all relevant paragraphs to find key evidence. In this paper, we employ two approaches for key evidence selection: a relevance score-based approach and a context-aware approach.

Relevance score-based selection methods rely on vector representations of statements and sentences in paragraphs. For a given statement $C$, we select sentences $s_i$ from the retrieved relevant passages $P = \{s_1, s_2, \ldots, s_n\}$ whose relevance score $r(c, s_i)$ is greater than a certain threshold $\lambda$ set experimentally. Details on setting lambda values can be found in Appendix A.2.3.

The context-aware sentence selection method uses a BERT-based sequence tagging model. We take as input the concatenation of statement claim $C = \{c_1, c_2, ..., c_k\}$ and passages $P = \{p_1, p_2, ..., p_m\}$ and separate them using special tokens: $[CLS]C[SEP]P[EOS]$. For the output of the model, we adopt the BIO token format, which classifies all irrelevant tokens as O, the first token of an evidence sentence as B evidence, and the remaining tokens of an evidence sentence as I evidence. We train a RoBERTa-large based model ~\cite{Liu2019RoBERTaAR}, minimizing the cross-entropy loss:
\begin{equation}
\mathcal{L}_{\theta} = -\sum_{i=1}^{N}\sum_{j=1}^{l_i}log(p\theta({y}_i^j))
\end{equation}
where $N$ is the number of examples in the training batch, $l_i$ is the number of non-padding tokens of the $i$-th example, and $p\theta({y}_i^j)$ is the estimated softmax probability of the correct label for the $j$-th token of the $i$-th example. We train this model on Factual-NLI~\cite{Factual-NLI} with batch size $64$, Adam optimizer, and initial learning rate  $5\times10^{-5}$ until convergence.


\textbf{Multi-step retrieval}: In the process of selecting key evidence, we assess the sufficiency of the evidence's relevance using a threshold $\lambda$. When the evidence is insufficient, we use iterative retrieval to supplement information. To prioritize the most significant fragments in the paragraph, we rank the selected fragments based on their scores. Similar to human behavior of recursively querying external sources like Wikipedia step by step until the desired information is found, only the fragments with the highest scores will be kept. The fragment with the highest score, referred to as the "winner," is then incorporated into the current query $[C[SEP]snippet]$. A reformulated query will be generated by combining the current query with current relevant paragraph information and updating it through a transformer.

\begin{equation}
C_{i+1} = Transformer([C_i [SEP]snippet])
\end{equation}
The reformulated query is fed back to the retriever, which uses it to reformulate and rank the passages in the corpus. $C_i$ fully interacts with the snippet through the transformer, avoiding information loss during the embedding process. The new query $C_{i+1}$ is again subjected to paragraph retrieval and key evidence selection, achieving the effect of multi-step iterative retrieval. This multi-step iterative approach allows our model to combine the multi-step information needed to validate claims from multiple Wikipedia pages.

\begin{table}[t]
  \caption{Statistics of three datasets.}
  \label{tab:datasets}
  \begin{tabular}{lccc}
    \toprule
     \multicolumn{1}{l}{\textbf{Platform}} & \multicolumn{1}{c}{\textbf{PolitiFact}} & \multicolumn{1}{c}{\textbf{GossipCop}} & \multicolumn{1}{c}{\textbf{Weibo}} \\
    \midrule
    \#Real News & 399 & 4,219 & 436\\
    \#Fake News & 345 & 3,393 & 311\\
    \#Total & 744 & 7,612 & 747\\
  \bottomrule
\end{tabular}
\end{table}

\subsection{Text Reasoning Module}
The last step of our model is to infer whether the news is true or false through multi-step retrieved evidence and news statements. This step aligns with human behavior, where individuals gather information from external sources and then evaluate the credibility of the news based on that information. Given a news claim $C$ and relevant evidence $E$ retrieved through a multi-step retrieval process, our text reasoning module performs a logical inference from the evidence to the claim. The textual reasoning model acts as an evaluator to judge whether a statement is logically consistent with the retrieved evidence, thus identifying a pair of claims and related evidence as true or false. Thus, the training task of a text reasoning model can be perceived as a binary classification task, where the goal is to minimize the binary cross-entropy loss function for each news item and its associated evidence. The cross-entropy loss is defined as follows:
\begin{equation}
\mathcal{L}_{CE} = -\frac{1}{N}\sum_{i=1}^{N}y_ilog(V(C_i,E_i)) + (1-y_i)log(1-V(C_i,E_i))
\end{equation}
$N$ is the number of samples in the current batch, $y = 1$ means that claim $C$ and evidence $E$ are logically consistent, and $y = 0$ means that $C$ and $E$ are contradictory. $V$ is a pre-trained language model that can perform discriminative classification tasks, such as BERT~\cite{bert2019}, ALBERT~\cite{albert2020}  and RoBERTa~\cite{Liu2019RoBERTaAR}.
In this work we choose BERT as the discriminator, we concatenate the claim $C$ and the evidence $E$ as the input of the discriminator, the input is [CLS] {C} [SEP] {E} [SEP], the batch size $N$ is 64, Adam optimizer and an initial learning rate of $5\times10^{-5}$ until convergence.

\section{EXPERIMENTS}
To verify the effectiveness of our proposed model, we conduct extensive experimental studies on three real-world datasets. Four research questions are addressed through comprehensive experimentation:

\begin{itemize}
\item RQ1: Is our MUSER model able to achieve improved fake news detection performance compared to previous fake news detection baseline methods?
\item RQ2: How does the impact of the number of steps in multi-step retrieval on model performance?
\item RQ3: How does each module of the model contribute to improved fake news detection performance?
\item RQ4: Is the evidence retrieved by our model meaningful and explainable through multi-step retrieval?
\end{itemize}

\subsection{Experimental Setup}
\subsubsection{Datasets.}We conduct experiments on three real-world datasets for fake news detection, including two English datasets (PolitiFact and GossipCop) and one Chinese dataset (Weibo). The English datasets PolitiFact and GossipCop are collected through FakeNewsNet~\cite{shu2020fakenewsnet}. The Weibo dataset is obtained through crawler tools~\cite{WuShu-weibo}. Their key statistics are shown in Table \ref{tab:datasets}.

\textbf{PolitiFact}: Within this dataset, the news articles are divided into two distinct categories: real news and fake news. This classification is determined based on the assessments provided by journalists and experts who review political news on various websites.

\textbf{GossipCop}: In this dataset, entertainment news articles with ratings are collected from various media.

\textbf{Weibo}: The data in this dataset are hot news topics from the Sina Weibo platform, and news is marked as rumors and non-rumors.

The datasets mentioned above contain both labeled news content and associated social information. However, since our work centers on curbing the initial propagation of fake news, we only utilize the news text without social information. This scenario resembles the situations where fake news detection must be performed before social information becomes available.

\subsubsection{Baselines} We compare MUSER with several existing methods, including content-based and evidence-based verification, as described below:

\textbf{Content-based methods}

\begin{itemize}
\item \textbf{TextCNN (EMNLP'14)}~\cite{kim-2014-textCNN}: TextCNN combines convolutional neural networks and news content, which can automatically extract text features through multiple convolutional hidden layers, 
\item \textbf{TextRNN (ACL'16)}~\cite{yang-etal-2016-textRNN}: TextRNN uses LSTM to encode the textual information in the last output of the recurrent neural network.
\item \textbf{TCNNURG (IJCAI'18))}~\cite{tcnn-urg}: TCNNURG utilizes two convolutional neural networks and a conditional variational autoencoder for classification.
\item \textbf{BERT (NAACL'19)}~\cite{bert2019}: BERT uses the Transformer-based architecture to pre-train deep bidirectional representations of unlabeled text.
\end{itemize}

\textbf{Evidence-based methods}
\begin{itemize}
\item \textbf{DeClarE (EMNLP'18)}~\cite{2018-declare}: They use BiLSTM to embed the semantics of evidence and compute evidence scores through an attention interaction mechanism.
\item \textbf{HAN (ACL'19)}~\cite{2019-han}: HAN adopts GRU embedding and two modules of topic consistency and semantic entailment based on a sentence-level attention mechanism to simulate claim-evidence interaction.
\item \textbf{EHIAN (IJCAI'20)}~\cite{2020-ehian}: EHIAN discusses the questionable parts of claims for interpretable claim verification through an evidence-aware hierarchical interactive attention network to explore more plausible evidence semantics.
\item \textbf{MAC (ACL'21)}~\cite{2021-mac}: MAC combines multi-head word-level attention and multi-head document-level attention, which facilitates interpretation for fake news detection at both word-level and evidence-level.
\item \textbf{GET (WWW'22)}~\cite{2022-get}: GET models claims and pieces of evidence as graph-structured data to explore complex semantic structures and reduces information redundancy through the semantic structure refinement layer. 
\end{itemize}

\subsubsection{Implementation Details} Fake news detection is commonly perceived as a binary classification problem, and the indicators used for model performance evaluation are F1, Precision, Recall, F1-Macro, and F1-Micro ~\cite{2022-get}. The dataset is partitioned into two sets, with 75\% of the data as the training set and the remaining 25\% of the data as the test set. The learning rate of the Adam optimizer is uniformly set to $5\times10^{-5}$ across all datasets. And the number of training epochs is set to $20$ for both our model and the baselines. The hyperparameters for the baselines are configured based on corresponding papers, with key hyperparameters carefully tuned for optimal performance (e.g., learning rate and embedding size). All experiments are conducted on Linux servers equipped with GeForce RTX 3080 GPUs (32GB memory each) using PyTorch 1.8.0. The implementation details are in the appendix and repository.

\begin{table}\footnotesize
\caption{Performance comparison of Our model w.r.t. baselines. We repeat the experiment 10 times, and average the results. "F1-Ma" and "F1-Mi" denote the metrics F1-Macro and F1-Micro, respectively. "-T" represents "True News as Positive" and "-F" denotes "Fake news as Positive" in the context of computing the precision and recall values. 
A t-test is performed on five dataset splits, with $P < .05$. The superior outcomes are indicated in bold and statistically significant improvements are denoted by *. }
\label{tab:performance}
\centering
\tabcolsep=3pt
\begin{tabular}{p{30pt}<{\centering} p{20pt}<{\centering} p{20pt}<{\centering} p{18pt}<{\centering} p{18pt}<{\centering} p{18pt}<{\centering} p{18pt}<{\centering} p{18pt}<{\centering} p{18pt}<{\centering} }
\toprule[1pt]
\multirow{2}{*}{\textbf{Method}} & \multicolumn{8}{c}{\textbf{PolitiFact}}  \\ \cline{2-9}
 & \textbf{F1-Ma} & \textbf{F1-Mi} & \textbf{F1-T} &  \textbf{P-T} & \textbf{R-T} & \textbf{F1-F} & \textbf{P-F} & \textbf{R-F}   \\ \hline
 \textbf{TextCNN}  & 0.601  & 0.602  & 0.608  & 0.641  & 0.579  & 0.594  & 0.564  & 0.615   \\
 \textbf{TextRNN}  & 0.610  & 0.609  & 0.616  & 0.650  & 0.586  & 0.603  & 0.572  & 0.636   \\
 \textbf{TextURG}  & 0.621  & 0.619  & 0.637  & 0.651  & 0.624  & 0.601  & 0.587  & 0.617   \\
 \textbf{BERT}  & 0.597   & 0.598  & 0.608  & 0.619  & 0.599  & 0.586  & 0.577  & 0.597  \\
 \textbf{DeClarE}  & 0.654  & 0.651  & 0.656  & 0.689  & 0.673  & 0.651  & 0.613  & 0.664  \\
 \textbf{HAN}  & 0.661  & 0.660  & 0.679  & 0.676  & 0.682  & 0.643  & 0.650  & 0.637   \\
 \textbf{EHIAN}  & 0.664  & 0.663  & 0.674  & 0.680  & 0.651  & 0.650  & 0.628  & 0.627   \\
 \textbf{MAC}  & 0.678  & 0.675  & 0.700  & 0.695  & 0.704  & 0.653  & 0.655  & 0.645   \\
 \textbf{GET}  & 0.694  & 0.692  & 0.725  & 0.712  & 0.770  & 0.669  & 0.720  & 0.665  \\

\hline

\textbf{MUSER}      & \textbf{0.732*} & \textbf{0.729*} &\textbf{0.757*} & \textbf{0.735*} & \textbf{0.780*} &  \textbf{0.702*} & \textbf{0.728*} & \textbf{0.681*} \\
\bottomrule[1pt]
\end{tabular}
\end{table}

\begin{table}\footnotesize
\caption{Performance comparison of on GossipCop.}
\label{tab:performance2}
\centering
\tabcolsep=3pt
\begin{tabular}{p{30pt}<{\centering} p{20pt}<{\centering} p{20pt}<{\centering} p{18pt}<{\centering} p{18pt}<{\centering} p{18pt}<{\centering} p{18pt}<{\centering} p{18pt}<{\centering} p{18pt}<{\centering} }
\toprule[1pt]
\multirow{2}{*}{\textbf{Method}} & \multicolumn{8}{c}{\textbf{GossipCop}}  \\ \cline{2-9}
 & \textbf{F1-Ma} & \textbf{F1-Mi} & \textbf{F1-T} &  \textbf{P-T} & \textbf{R-T} & \textbf{F1-F} & \textbf{P-F} & \textbf{R-F}   \\ \hline
 \textbf{TextCNN}  & 0.628  & 0.624  & 0.658  & 0.671  & 0.646  & 0.590  & 0.604  & 0.576   \\
 \textbf{TextRNN}  & 0.629  & 0.628  & 0.636  & 0.667  & 0.609  & 0.620  & 0.591  & 0.651   \\
 \textbf{TextURG}  & 0.644  & 0.643  & 0.650  & 0.684  & 0.619  & 0.636  & 0.605  & 0.637   \\
 \textbf{BERT}  & 0.617   & 0.613  & 0.635  & 0.664  & 0.649  & 0.578  & 0.635  & 0.562  \\
 \textbf{DeClarE}  & 0.660  & 0.657  & 0.686  & 0.677  & 0.694  & 0.629  & 0.638  & 0.619  \\
 \textbf{HAN}  & 0.702  & 0.700  & 0.722  & 0.721  & 0.716  & 0.678  & 0.676  & 0.680   \\
 \textbf{EHIAN}  & 0.705  & 0.702  & 0.731  & 0.713  & 0.749  & 0.673  & 0.694  & 0.654   \\
 \textbf{MAC}  & 0.729  & 0.727  & 0.725  & 0.742  & \textbf{0.756}  & 0.705  & 0.713  & 0.697   \\
\textbf{GET}  & 0.733  & 0.731  & 0.751  & 0.749  & 0.727  & 0.712  & 0.710  & 0.715  \\

\hline

\textbf{MUSER}     & \textbf{0.776*} & \textbf{0.775*} & \textbf{0.784*} & \textbf{0.843*} &0.734 &\textbf{0.768*} &\textbf{0.714*} &\textbf{0.830*} \\
\bottomrule[1pt]
\end{tabular}
\end{table}

\begin{table}\footnotesize
\caption{Performance comparison of on Weibo.}
\label{tab:performance3}
\centering
\tabcolsep=3pt
\begin{tabular}{p{30pt}<{\centering} p{20pt}<{\centering} p{20pt}<{\centering} p{18pt}<{\centering} p{18pt}<{\centering} p{18pt}<{\centering} p{18pt}<{\centering} p{18pt}<{\centering} p{18pt}<{\centering} }
\toprule[1pt]
\multirow{2}{*}{\textbf{Method}} & \multicolumn{8}{c}{\textbf{Weibo}}  \\ \cline{2-9}
 & \textbf{F1-Ma} & \textbf{F1-Mi} & \textbf{F1-T} &  \textbf{P-T} & \textbf{R-T} & \textbf{F1-F} & \textbf{P-F} & \textbf{R-F}   \\ \hline
 \textbf{TextCNN}  & 0.722  & 0.721  & 0.740  & 0.742  & 0.736  & 0.703  & 0.706 & 0.700   \\
 \textbf{TextRNN}  & 0.741  & 0.737  & 0.771  & 0.730  & 0.812  & 0.701  & 0.756  & 0.654   \\
 \textbf{TextURG}  & 0.709  & 0.704  & 0.741  & 0.712  & 0.628  & 0.667  & 0.707  & 0.759   \\
 \textbf{BERT}  & 0.699   & 0.698  & 0.719  & 0.720  & 0.716  & 0.678  & 0.676  & 0.680  \\
 \textbf{DeClarE}  & 0.746  & 0.745  & 0.765  & 0.758  & 0.771  & 0.724  & 0.732  & 0.717  \\
 \textbf{HAN}  & 0.689  & 0.687  & 0.711  & 0.706  & 0.716  & 0.662  & 0.668  & 0.657   \\
 \textbf{EHIAN}  & 0.753  & 0.752  & 0.770  & 0.768  & 0.772  & 0.734  & 0.754  & 0.731  \\ 
 \textbf{MAC}  & 0.734  & 0.732  & 0.709  & 0.722  & 0.697  & 0.755  & 0.745  & 0.766   \\
 \textbf{GET}  & 0.756  & 0.754  & 0.776  & 0.760  & 0.794  & 0.730  & 0.761  & 0.712  \\

\hline

\textbf{MUSER}      & \textbf{0.804*} & \textbf{0.802*} & \textbf{0.824*} & \textbf{0.812*} & \textbf{0.837*} & \textbf{0.791*} & \textbf{0.806*} & \textbf{0.778*} \\
\bottomrule[1pt]
\end{tabular}
\end{table}

\begin{figure*}[t]
  \includegraphics[width=0.85\textwidth]{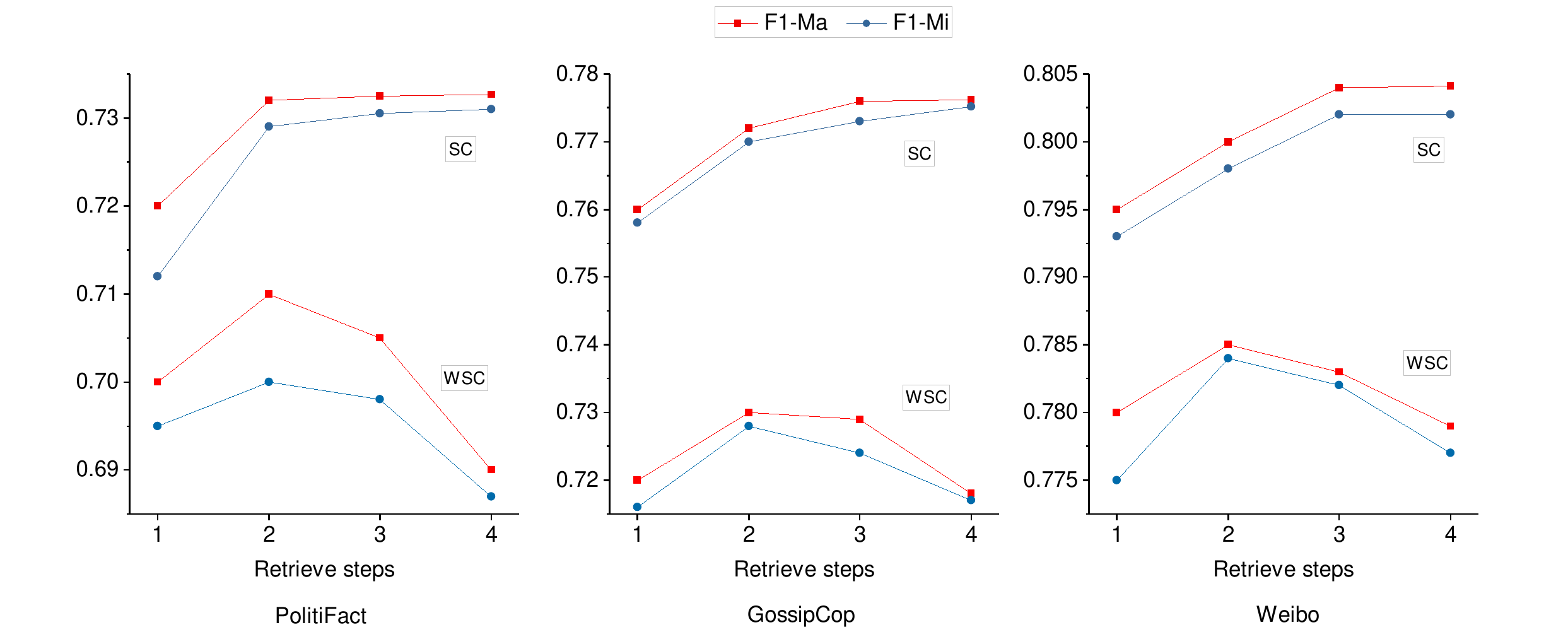}
  \caption{Results of retrieve step comparison study. The term SC (Step Control) means that the key evidence selection function is activated, while WSC (Without Step Control) means that the key evidence selection function is not included. }
  \Description{xxx}
  \label{step_control}
\end{figure*}

\subsection{Performance Results (RQ1)}
We compare our model, MUSER, to $9$ baselines, including $4$ content-based methods and $5$ evidence-based methods. The results are reported in Tables \ref{tab:performance}, \ref{tab:performance2}, and \ref{tab:performance3}, and we have the following observations:

Firstly, it is worth noting that evidence-based methods tend to predict more correctly than content-based methods (i.e., the first four methods in the tables), indicating the extra value of incorporating additional evidential information, which can well make up for the insufficiency of news content features alone. The evidence-based methods rely on external evidence to verify the validity of the claims, reducing excessive reliance on textual schemas.

Secondly, in comparison to three recent evidence-based methods (GET, EHIAN, MAC), our proposed MUSER achieves superior results (MUSER > GET > EHIAN > MAC). In particular, MUSER improves the performance by $3\%$ on F1-Macro and F1-Micro compared to the current SOTA baseline GET on the three datasets, which can better reflect the overall detection ability of the model. Furthermore, for more fine-grained evaluation, we computed "True news as Positive" and "Fake news as Positive" separately. MUSER also achieved superior results in F1, Precision, and Recall scores on the three datasets. Accuracy is equivalent to F1-Macro and thus omitted in the evaluation.

Finally, our results demonstrate that MUSER outperforms all baseline methods in fake news detection, as indicated by the positive detection metric. For instance, as far as GossipCop is concerned, the F1-False, Precision-False, and Recall-False values have been increased by 5\%, 0.4\%, and 11\%, respectively. Similar obvious improvements can be observed on other datasets. These results show that our method exhibits a higher degree of accuracy in discerning fake news. 
Enhanced by multi-step iterative evidence retrieval, our model can extract relevant information, so as to better assist in assessing the veracity of news. 
Furthermore, extensive experiments are conducted on large public datasets for the detection of fake news. Detailed information can be found in Appendix A.2.1.


\begin{figure}[t]
  \centering
  \includegraphics[width=\linewidth]{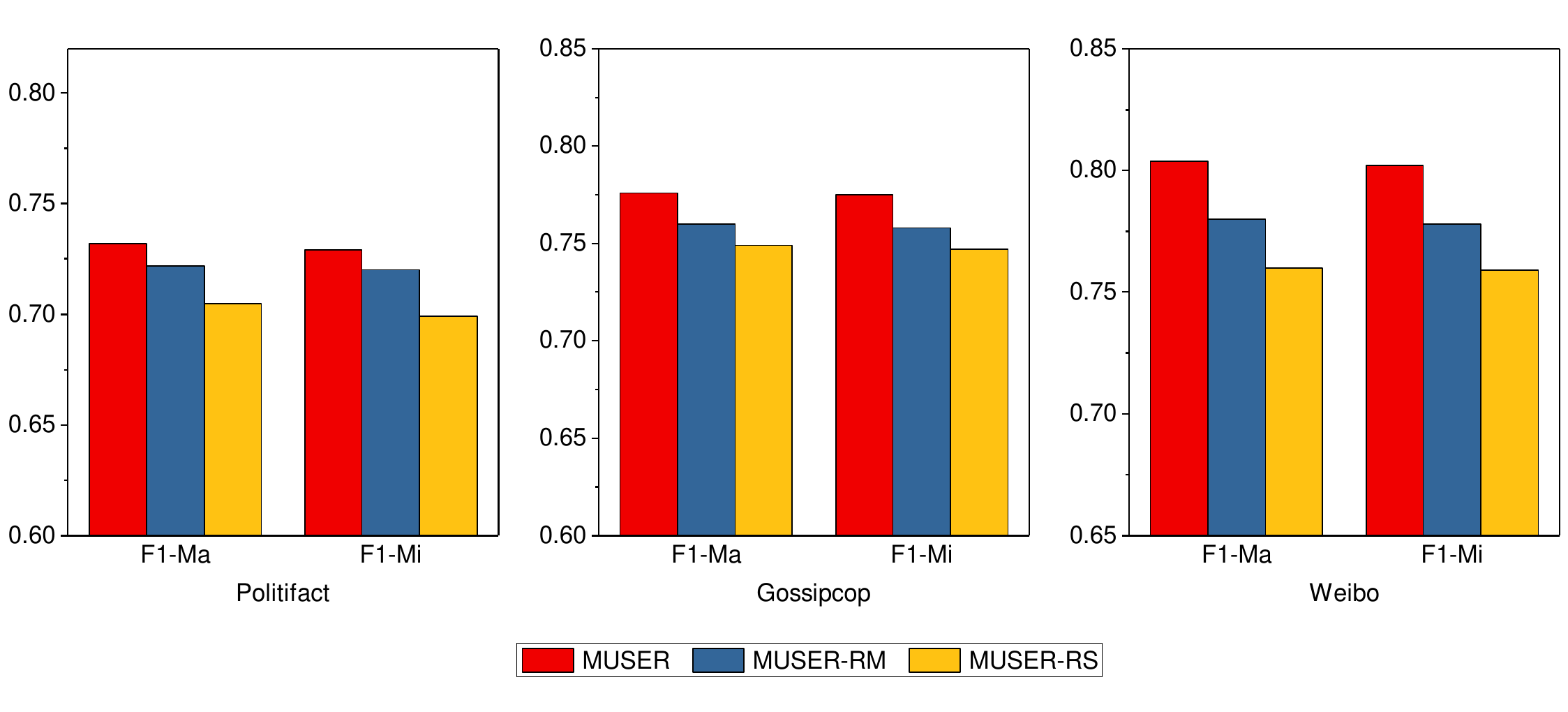}
  \caption{Results of ablation study. MUSER represents the complete model performance, MUSER-RM represents the removal of the multi-step retrieval module and MUSER-RS represents the removal of the text summary module.}
  \Description{xxx}
  \label{ablation}
\end{figure}

\subsection{Retrieve Steps Comparison (RQ2)}
Next, we investigate the performance improvement of the number of retrieval steps in the multi-step retrieval module. The evaluation is conducted using the commonly used F1-Macro and F1-Micro scores on each dataset and results are presented in Figure \ref{step_control}. In order to examine the effectiveness of key evidence selection in the multi-step retrieval process, we remove it and use a fixed number of retrieval steps to conduct experiments, and then compare it to the model with the key evidence selection function.

Firstly, we can find that in experiments where key evidence selection is not enabled, as the number of retrievals increases, the performance decreases instead. This is because there is no evidence screening for the retrieved paragraphs, which may contain redundant information, leading to a decrease in performance.

Secondly, we observe that enabling key evidence selection results in improved performance compared to the scenario where key evidence selection is not enabled. In the key evidence selection stage, our model determines whether the current retrieval results include key evidence. When key evidence is successfully retrieved, the iterative retrieval process is halted to minimize the interference caused by redundant information. In other words, the selection strategy follows an exploratory approach, where the emphasis is on exploring relevant information first. Importantly, increasing the number of retrieval steps does not result in an increase in redundant information.

The key takeaway from this experiment is that multiple retrieval steps consistently improve performance compared to single-step retrieval. That is, even if relevant evidence passages are not retrieved in the initial step, the retriever will continue in the subsequent iterative retrieval process. 
The performance of the model reaches its peak around 2 to 3 retrieval steps. Beyond this point, increasing the number of steps does not yield significant benefits and, in fact, leads to a degradation in performance.
Interestingly, despite variations in the difficulty level of the datasets, the optimal number of retrieval steps remains consistent.

\subsection{Ablation Study (RQ3)}
In this part, comparative performance experiments are conducted to assess the necessity of each module. 
As depicted in Figure \ref{ablation}, MUSER outperforms MUSER-RM, proving the critical role of multi-step iterative evidence retrieval. Additionally, the text summarization module is also important. By extracting key statements in the news, the interference of unrelated information is mitigated, thereby achieving more accurate predictions. Furthermore, MUSER performs better than MUSER-RS and MUSER-RM, showing that removing any of them leads to performance degradation, which demonstrates the effectiveness of our main components.

\subsection{Explainability Study (RQ4)}

\begin{figure}[t]
  \centering
  \includegraphics[width=\linewidth]{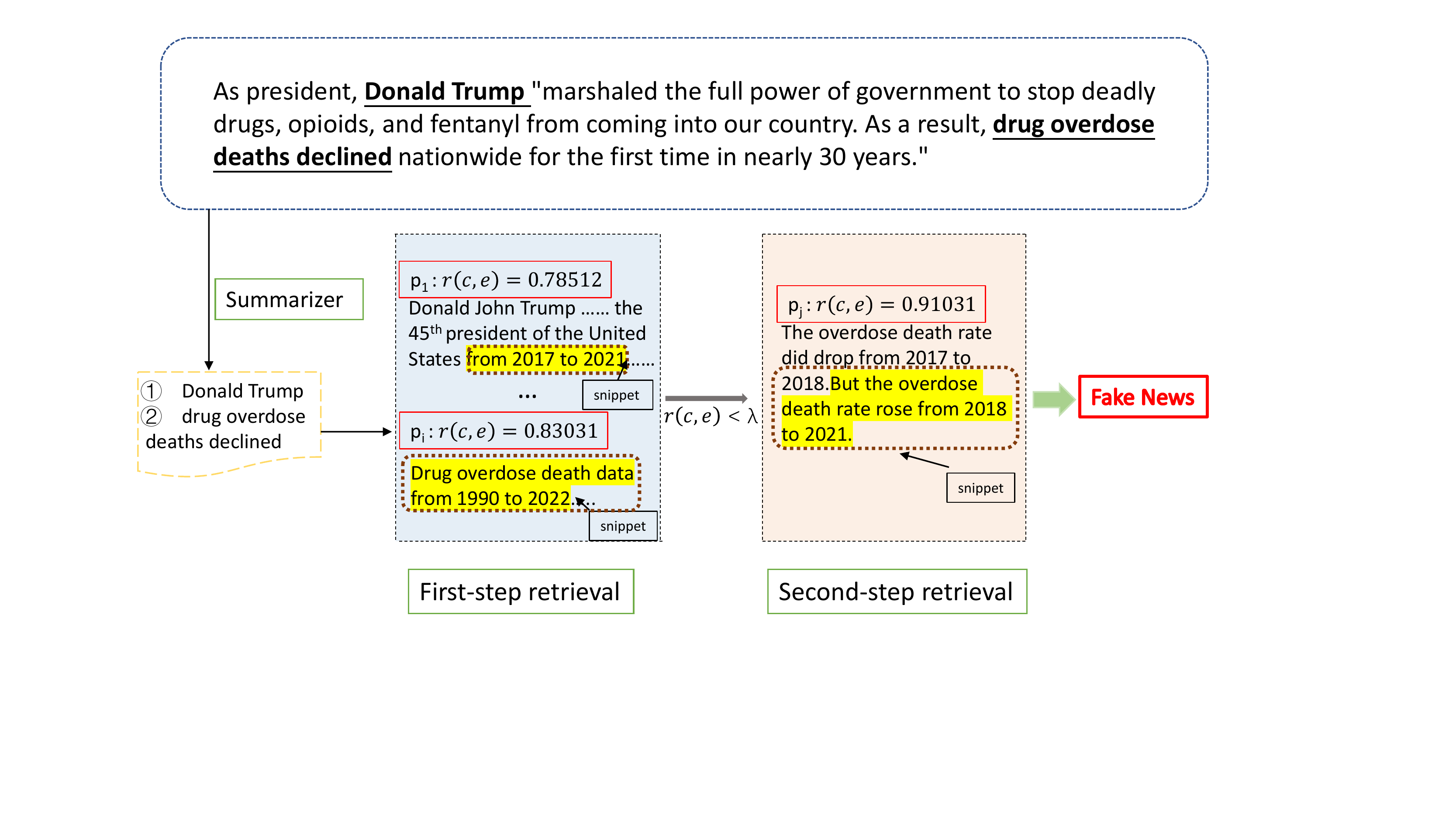}
  \caption{A verification example generated by MUSER in the Case study. The evidence correlation score $r(c,e)$ obtained by the first step of retrieval is smaller than the threshold $\lambda$ we set. Then proceed to the second step of retrieval to obtain more sufficient evidence.}
  \Description{xxx}
  \label{casestudy}
\end{figure}

\subsubsection{Case Study} 
In this part, we demonstrate the effectiveness of our model in facilitating a deeper understanding of the multi-step retrieval process. In particular, we present a specific example involving the evaluation of a news story concerning US President Donald Trump's efforts to combat drug-related issues. The news says "Donald Trump marshaled the full power of government to stop deadly drugs, opioids, and fentanyl from coming into our country. As a result, drug overdose deaths declined nationwide for the first time in nearly 30 years." By employing key evidence extraction and conducting a multi-step search for supplementary evidence, MUSER successfully identifies this news as fake. This particular case serves as a compelling demonstration of MUSER's capability to accurately assess the authenticity of the news. 

Specifically, Figure \ref{casestudy} shows the steps of the verification processes. After key information is extracted in the text summarization model, the first step of retrieval is performed, and relevant paragraph data is obtained from the corpus. Evidence extraction identifies information related to Donald Trump and data on drug overdose deaths in the United States. The calculated $r(c,e)$ from the key evidence selection is less than the preset limit value $\lambda$, indicating the necessity for another retrieval step. In the second step, the snippet information retrieved is carried forward and the statement "The overdose death rate did drop from 2017 to 2018. But the overdose death rate rose from 2018 to 2021." is obtained. Finally, the reasoning module judges the news to be fake. 
Evidence from multi-step retrieval makes it easier for users to understand the judgments made by the model on the authenticity of the news.

\subsubsection{User Study} In this part, we aim to determine if real-world users are able to accurately assess the veracity of news articles based on the evidence retrieved by MUSER. Specifically, we conduct a user study in which there are $60$ news articles randomly selected from PolitiFact, GossipCop, and Weibo, with $10$ fake and $10$ real news articles from each dataset.
We compare the evidence retrieved by MUSER with the evidence obtained by the GET model after refinement by semantic structure and ask $8$ participants to score the evidence. 
For each piece of news, we will give the relevant evidence of MUSER or GET, and then ask the participant to determine whether the news is true or fake based on the given evidence within three minutes. Moreover, participants are asked to give an adjusted confidence score about her/his conclusion according to a 5-point Likert scale.
To ensure fairness in our user study experiment, each participant is given the news articles to be judged in a randomized manner and participate in the experiment independently. 

Table \ref{tab:userstudy} shows the results of the experiments. By comparing the labels given by different participants, we find that the conclusions drawn by the participants have a high level of consistency with the predicted labels produced by the MUSER model.
This indicates that by observing the multi-step retrieval of evidence generated by MUSER, human participants can much more accurately decide whether a news article is fake or not.

\begin{table}[t]
  \caption{Results of the user study. The agreement measure means the proportion of concurrence between the user's judgment and the model's judgment.}
  \label{tab:userstudy}
\begin{tabular}{|c|c|c|c|c|} \hline %
\textbf{Method} & \textbf{F1} & \textbf{Precision} & \textbf{Agreement} \\ \hline
\textbf{GET} & 0.690 & 0.667 & 70\%   \\ \hline
\textbf{MUSER} &	\textbf{0.758} & \textbf{0.733} & \textbf{76.7\%}  \\ \hline
\end{tabular}
\end{table}

\section{Conclusion}

In this paper, we propose a framework for fake news detection based on multi-step evidence retrieval enhancement---MUSER. Our model leverages a three-phase methodology inspired by human verification processes, including summarization, retrieval, and reasoning. Through text summarization, key information is extracted from the news, reducing irrelevant information. The multi-step retrieval phase enables evidence association for news verification, increasing the dependency between multiple pieces of evidence. Finally, the semantic connection between the news statement and the evidence is analyzed for news classification into two categories: true news and fake news. The results of our experiments on three real-world demonstrated its effectiveness. Moreover, our results also show that evidence association via multi-step retrieval enhances the interpretability of the fake news detection task, making it easier for users to assess the credibility of information and form their own valid judgments. 

\section{Acknowledgments}
\begin{sloppypar}
Thanks to Dr. Shu Wu, Dr. Xiting Wang, Dr. Jianxun Lian, and the anonymous reviewers for their valuable comments and constructive feedback. This work is supported by the National Natural Science Foundation of China (Grant Nos. 62276171 and 62072311), Guangdong Provincial Key Laboratory of Popular High Performance Computers, Shenzhen Fundamental Research-General Project (Grant Nos. JCYJ20190808162601658, 20220811155803001, 20210324094402008 and 20200814105901001), CCF-Baidu Open Fund (Grant No. OF2022028), and Swiftlet Fund Fintech funding. Kai Shu is supported by NSF SaTC-2241068. Kai Shu and Xing Xie are the corresponding authors.
\end{sloppypar}

\bibliographystyle{unsrt}


\appendix

\section{Appendix On Reproducibility}

\setcounter{table}{0}
\setcounter{figure}{0}
\renewcommand{\thetable}{A\arabic{table}}
\renewcommand{\thefigure}{A\arabic{figure}}

\subsection{Experimental Environment}
The experiments are conducted on CentOS 7 servers equipped with GeForce RTX 3080 GPUs, each with 32GB of memory. The code for the experiment is implemented using PyTorch version 1.8.0.

\subsection{Supplementary Experiment}
\subsubsection{Large Dataset Experiments}
We further validate the performance of MUSER on two large public datasets. The first is the LIAR dataset (\href{https://www.cs.ucsb.edu/~william/data/liar_dataset.zip}{https://www.cs.ucsb.e du/~william/data/liar\_dataset.zip}). We transform the original LIAR multi-class dataset into a binary classification format, where each sample is labeled as either "true" or "false". We merge the original multiple categories into these two binary labels. We merge "mostly true" and "true" into "true", and "barely-true", "false", and "pants-fire" into "false" to better suit the needs of binary classification problems. Moreover, we compare two representative baseline methods, and the experimental results are given in Table \ref{tab:performance-LIAR}.


\begin{table}\footnotesize
\caption{Performance comparison of on LIAR.}
\label{tab:performance-LIAR}
\centering
\tabcolsep=3pt
\begin{tabular}{p{30pt}<{\centering} p{20pt}<{\centering} p{20pt}<{\centering} p{18pt}<{\centering} p{18pt}<{\centering} p{18pt}<{\centering} p{18pt}<{\centering} p{18pt}<{\centering} p{20pt}<{\centering} }
\toprule[1pt]
\multirow{2}{*}{\textbf{Method}} & \multicolumn{8}{c}{\textbf{LIAR}}  \\ \cline{2-9}
 & \textbf{F1-Ma} & \textbf{F1-Mi} & \textbf{F1-T} &  \textbf{P-T} & \textbf{R-T} & \textbf{F1-F} & \textbf{P-F} & \textbf{R-F}   \\ \hline
\textbf{BERT}  & 0.57142   & 0.54946  & 0.51000  & 0.63879  & 0.55120  & 0.58893  & 0.50217  & 0.62307  \\
\textbf{GET}  & 0.61413  & 0.61048  & 0.57280  & 0.56667  & 0.57907  & 0.64116  & \textbf{0.65400}  & 0.63243  \\
\hline

\textbf{MUSER}     & \textbf{0.64502} & \textbf{0.64413} & \textbf{0.64731} & \textbf{0.64042} &\textbf{0.65434} &\textbf{0.64270} &0.64977 &\textbf{0.63577} \\
\bottomrule[1pt]
\end{tabular}
\end{table}

We also conduct experiments on the Fakeddit dataset (\href{https://github.com/entitize/Fakeddit}{https://gith ub.com/entitize/Fakeddit}), which contains multi-modal fake news data (image, text). For this experiment, we select textual data only. The label used is the '2-way' introduced in the data set, which is divided into two categories: true and false. The experimental results are given in Table \ref{tab:performance-Fakeddit}.


\begin{table}\footnotesize
\caption{Performance comparison of on Fakeddit.}
\label{tab:performance-Fakeddit}
\centering
\tabcolsep=3pt
\begin{tabular}{p{30pt}<{\centering} p{20pt}<{\centering} p{20pt}<{\centering} p{18pt}<{\centering} p{18pt}<{\centering} p{18pt}<{\centering} p{18pt}<{\centering} p{18pt}<{\centering} p{20pt}<{\centering} }
\toprule[1pt]
\multirow{2}{*}{\textbf{Method}} & \multicolumn{8}{c}{\textbf{Fakeddit}}  \\ \cline{2-9}
 & \textbf{F1-Ma} & \textbf{F1-Mi} & \textbf{F1-T} &  \textbf{P-T} & \textbf{R-T} & \textbf{F1-F} & \textbf{P-F} & \textbf{R-F}   \\ \hline
\textbf{BERT}  & 0.82086 &	0.75175 &	0.88273 &	0.87511 &	0.89048 &	0.62076 &	0.63874 &	0.60377  \\
\textbf{GET}  & 0.84651 &	0.79794 &	0.89700 &	0.87060 &	0.92506 &	0.68538 &	0.76689 &	0.61195  \\
\hline

\textbf{MUSER}     & \textbf{0.86785} & \textbf{0.82201} & \textbf{0.91611} & \textbf{0.88847} &\textbf{0.94552} &\textbf{0.68887} &\textbf{0.77871} &\textbf{0.62761} \\
\bottomrule[1pt]
\end{tabular}
\end{table}

\subsubsection{Retrieval Budget Experiments}
In order to assess the effectiveness of multi-step retrieval and investigate whether increasing the retrieval budget can serve as a substitute, we performed the following experiments.
To compare with our MUSER's configuration, which is a 3-step retrieval with $N1 = 30$, $N2 = 30$, and $N3 = 30$, we use three configurations of 1-step retrieval with $N1 = 30$, $N1 = 60$ or $N1 = 90$. The results are reported in Table \ref{tab:performance-budget}. Surprisingly, increasing the pool of 1-step retrieval has a negative effect on the performance, and the patterns are consistent across the three datasets.
The reasons for this can be two folds. First, there exists a degree of interdependence among certain pieces of evidence, which requires additional information from previous retrieval steps for accurate identification. Second, the inclusion of an excessive number of paragraphs introduces a higher level of noise into the text, ultimately leading to suboptimal results. Consequently, our investigation has demonstrated that the proposed 1-step retrieval with an augmented budget is not a viable alternative to MUSER's multi-step retrieval approach.

\begin{table}\footnotesize
\caption{Performance comparison of retrieval budget experiments. }
\label{tab:performance-budget}
\centering
\tabcolsep=3pt
\begin{tabular}{p{30pt}<{\centering} p{28pt}<{\centering} p{28pt}<{\centering} p{28pt}<{\centering} p{28pt}<{\centering} p{28pt}<{\centering} p{26pt}<{\centering}}
\toprule[1pt]
\multirow{2}{*}{\textbf{Method}} & \multicolumn{2}{c}{\textbf{PolitiFact}} & \multicolumn{2}{c}{\textbf{Gossipcop}} & \multicolumn{2}{c}{\textbf{Weibo}} \\ \cline{2-7}
 & \textbf{F1-Ma} & \textbf{F1-Mi} & \textbf{F1-Ma} & \textbf{F1-Mi} & \textbf{F1-Ma} & \textbf{F1-Mi}   \\ \hline
\textbf{N1 = 30}  & 0.7193	&0.7155	&0.7600	&0.7580	&0.7951	&0.7930    \\
\textbf{N1 = 60}  &  0.7117	&0.7110	&0.7560	&0.7555	&0.7868	&0.7859\\
\textbf{N1 = 90}  &  0.7064	&0.7057	&0.7546	&0.7542	&0.7855	&0.7846  \\
\hline

\textbf{MUSER}     & \textbf{0.7324} & \textbf{0.7293} & \textbf{0.7760} & \textbf{0.7751} &\textbf{0.8042} &\textbf{0.8026}  \\
\bottomrule[1pt]
\end{tabular}
\end{table}

\balance

\subsubsection{Threshold $\lambda$ selection}
$\lambda$ essentially serves as an evaluation metric for the correlation between evidence and passage text. In our research, we employed a unified threshold value of $\lambda = 0.9$ across all datasets. To investigate the effect of $\lambda$ value, we conduct experiments to analyze its impact on the retrieved evidence. And the results reveal that a low $\lambda$ value tends to introduce more noise in the retrieved evidence, whereas a high $\lambda$ value may inadvertently exclude critical evidence. For your reference, the comparison results with different $\lambda$ values are given in Table \ref{tab:performance-lambda}.

\begin{table}\footnotesize
\caption{The comparison results with different $\lambda$ values. }
\label{tab:performance-lambda}
\centering
\tabcolsep=3pt
\begin{tabular}{p{30pt}<{\centering} p{28pt}<{\centering} p{28pt}<{\centering} p{28pt}<{\centering} p{28pt}<{\centering} p{28pt}<{\centering} p{26pt}<{\centering}}
\toprule[1pt]
\multirow{2}{*}{\textbf{ $\lambda$ }} & \multicolumn{2}{c}{\textbf{PolitiFact}} & \multicolumn{2}{c}{\textbf{Gossipcop}} & \multicolumn{2}{c}{\textbf{Weibo}} \\ \cline{2-7}
 & \textbf{F1-Ma} & \textbf{F1-Mi} & \textbf{F1-Ma} & \textbf{F1-Mi} & \textbf{F1-Ma} & \textbf{F1-Mi}   \\ \hline
\textbf{0.8}  & 0.699	&0.698	&0.734	&0.735	&0.787	&0.786    \\
\textbf{0.85}  & 0.712	&0.710	&0.755	&0.753	&0.796	&0.795\\

\textbf{0.9}     & \textbf{0.732} & \textbf{0.729} & \textbf{0.776} & \textbf{0.775} &\textbf{0.804} &\textbf{0.802}  \\

\textbf{0.9}  &  0.715	&0.713	&0.739	&0.738	&0.789	&0.788
  \\

\bottomrule[1pt]
\end{tabular}
\end{table}

\subsection{Code Resources}

We compare the proposed framework, MUSER, with 9 baseline methods discussed in Section 5.2, the content-based methods including TextCNN, TextRNN, TCNNURG, BERT, and the evidence-based methods including DeClarE, HAN, EHIAN, MAC, and GET. The implementation details of our proposed framework, including code and settings, are available through the following link: ~\href{https://github.com/Complex-data/MUSER}{https://github. com/Complex-data/MUSER}. Other codes were obtained as follows:
\begin{itemize}
    \setlength{\itemsep}{0pt}
    \setlength{\parsep}{0pt}
    \setlength{\parskip}{0pt}
    
    \item \textbf{TextCNN:} we use the publicly available implementation at: \href{https://github.com/FinIoT/text_cnn}{https://github.com/FinIoT/text\_cnn}
    
    \item \textbf{TextRNN:} we use the publicly available implementation at: \href{https://github.com/luchi007/RNN\_Text\_Classify}{https://github.com/luchi007/ RNN\_Text\_Classify}

    \item \textbf{TCNNURG:} we use the publicly available implementation at: \href{https://github.com/ttjjlw/text_classify_by_keras}{https://github.com/text\_classify}
    
    \item \textbf{BERT:} we use the publicly available implementation at: \href{https://github.com/google-research/bert}{https://github.com/google-research/bert}
    
    \item \textbf{DeClarE:} we use the publicly available implementation at: \href{https://github.com/atulkumarin/DeClare}{https://github.com/atulkumarin/DeClare}
    
    \item \textbf{HAN:} we use the publicly available implementation at: \href{https://github.com/majingCUHK/Claim_Verification}{https: //github.com/majingCUHK/Claim\_Verification}
    
    \item \textbf{EHIAN:} we use the publicly available implementation at: \href{https://github.com/jayded/evidence-inference}{https://github.com/evidence-inference}
    
    \item \textbf{MAC:} we use the publicly available implementation at: \href{https://github.com/nguyenvo09/EACL2021}{https: //github.com/nguyenvo09/EACL2021}

    \item \textbf{GET:} we use the publicly available implementation at: \href{https://github.com/CRIPAC-DIG/GET}{https: //github.com/CRIPAC-DIG/GET}
    
\end{itemize}

\subsection{Corpus processing}
In this article, we use Wikipedia data as the retrieval corpus. The download address of Wikipedia Chinese corpus is: \href{https://dumps.wikimedia.org/zhwiki/latest/}{https://dumps.  wikimedia.org/zhwiki/latest/}, and the download address of Wikipedia English corpus is: \href{https://dumps.wikimedia.org/enwiki/latest/}{https://dumps.wikimedia.org/enwiki/latest/}.

We extract the Wikipedia corpus through WikiExtractor, which can extract the main article content of the corpus ending with .bz downloaded from Wikipedia. The download address of the tool is: \href{https://github.com/attardi/wikiextractor}{https://github.com/attardi/wikiextractor}.

\end{document}